\documentclass[acmtog]{acmart}

\acmPrice{15.00}

\setcopyright{acmcopyright}
\acmJournal{TOG}
\acmYear{2019}
\acmVolume{38}
\acmNumber{4}
\acmArticle{72}
\acmMonth{7} 
\acmDOI{10.1145/3306346.3322966}

\usepackage{booktabs} 
\usepackage{xcolor}
\usepackage{amsmath}
\usepackage{amssymb}  
\usepackage{bm}
\usepackage{amsfonts}
\usepackage{multirow}
\usepackage{subcaption}
\setcitestyle{square}

\newcommand{\norm}[1]{\left\lVert#1\right\rVert}

\usepackage[ruled]{algorithm2e} 

\SetAlFnt{\small}
\SetAlCapFnt{\small}
\SetAlCapNameFnt{\small}
\SetAlCapHSkip{0pt}

\received{January 2019}
\received[final version]{April 2019}
\received[accepted]{April 2019}

\begin{document}
\title{Synthesis of Biologically Realistic Human Motion Using Joint Torque Actuation}

\author{Yifeng Jiang}
\affiliation{%
 \institution{Georgia Institute of Technology}
 \country{USA}
}
\email{yjiang340@gatech.edu}

\author{Tom Van Wouwe}
\affiliation{%
 \institution{KU Leuven}
 \country{Belgium}
}
\email{tom.vanwouwe@kuleuven.be}

\author{Friedl De Groote}
\affiliation{%
 \institution{KU Leuven}
 \country{Belgium}
}
\email{friedl.degroote@kuleuven.be}

\author{C. Karen Liu}
\affiliation{%
 \institution{Georgia Institute of Technology}
 \country{USA}
}
\email{karenliu@cc.gatech.edu}

\renewcommand\shortauthors{Jiang, Van Wouwe, De Groote, and Liu}

\begin{abstract}
Using joint actuators to drive the skeletal movements is a common practice in character animation, but the resultant torque patterns are often unnatural or infeasible for real humans to achieve. On the other hand, physiologically-based models explicitly simulate muscles and tendons and thus produce more human-like movements and torque patterns. This paper introduces a technique to transform an optimal control problem formulated in the muscle-actuation space to an equivalent problem in the joint-actuation space, such that the solutions to both problems have the same optimal value. By solving the equivalent problem in the joint-actuation space, we can generate human-like motions comparable to those generated by musculotendon models, while retaining the benefit of simple modeling and fast computation offered by joint-actuation models. Our method transforms constant bounds on muscle activations to nonlinear, state-dependent torque limits in the joint-actuation space. In addition, the metabolic energy function on muscle activations is transformed to a nonlinear function of joint torques, joint configuration and joint velocity. Our technique can also benefit policy optimization using deep reinforcement learning approach, by providing a more anatomically realistic action space for the agent to explore during the learning process. We take the advantage of the physiologically-based simulator, OpenSim, to provide training data for learning the torque limits and the metabolic energy function. Once trained, the same torque limits and the energy function can be applied to drastically different motor tasks formulated as either trajectory optimization or policy learning. 

\end{abstract}

%
%

%
%


\maketitle




\definecolor{darkblue}{rgb}{0.0, 0.0, 0.53}
\newcommand{\old}[1]{\textcolor{red}{\textbf {\st{#1}}}}
\newcommand{\new}[1]{\textcolor{red}{#1}}
\newcommand{\note}[1]{\cmt{Note: #1}}
\newcommand{\karen}[1]{\textcolor{red}{{Karen: #1}}}
\newcommand{\friedl}[1]{\textcolor{orange}{{Friedl: #1}}}
\newcommand{\tom}[1]{\textcolor{orange}{{Tom: #1}}}
\newcommand{\ck}[1]{\textcolor{Plum}{{Charlie: #1}}}
\newcommand{\original}[1]{\textcolor{blue}{{Original: #1}}}
\newcommand{\jie}[1]{\textcolor{blue}{{Jie: #1}}}
\newcommand{\alex}[1]{\textcolor{Purple}{{Alex: #1}}}
\newcommand{\greg}[1]{\textcolor{green}{{Greg: #1}}}
\newcommand{\zackory}[1]{\textcolor{Orange}{{Zackory: #1}}}
\newcommand{\charlie}[1]{\textcolor{cyan}{{Charlie: #1}}}
\newcommand{\henry}[1]{\textcolor{TealBlue}{{Henry: #1}}}
\newcommand{\wenhao}[1]
{\textcolor{ForestGreen}{{Wenhao: #1}}}
\newcommand{\yifeng}[1]
{\textcolor{darkblue}{{Yifeng: #1}}}
\newcommand{\newtext}[1]{#1}
\newcommand{\eqnref}[1]{Equation~(\ref{eqn:#1})}

\long\def\ignorethis#1{}

\newcommand{\etal}{{\em{et~al.}\ }}
\newcommand{\eg}{e.g.\ }
\newcommand{\ie}{i.e.\ }

\newcommand{\figtodo}[1]{\framebox[0.8\columnwidth]{\rule{0pt}{1in}#1}}
\newcommand{\figref}[1]{Figure~\ref{fig:#1}}
\newcommand{\secref}[1]{Section~\ref{sec:#1}}

\newcommand{\vc}[1]{\ensuremath{\boldsymbol{#1}}}
\newcommand{\pd}[2]{\ensuremath{\frac{\partial{#1}}{\partial{#2}}}}
\newcommand{\pdd}[3]{\ensuremath{\frac{\partial^2{#1}}{\partial{#2}\,\partial{#3}}}}

\newcommand{\vEndEff}{\ensuremath{\vc{d}}}
\newcommand{\vRelMove}{\ensuremath{\vc{r}}}
\newcommand{\sSet}{\ensuremath{S}}

\newcommand{\vControl}{\ensuremath{\vc{u}}}
\newcommand{\vPoint}{\ensuremath{\vc{p}}}
\newcommand{\sSpringCoef}{{\ensuremath{k_{s}}}}
\newcommand{\sDamperCoef}{{\ensuremath{k_{d}}}}
\newcommand{\vHandle}{\ensuremath{\vc{h}}}
\newcommand{\vForce}{\ensuremath{\vc{f}}}

\newcommand{\mTransChain}{\ensuremath{\vc{W}}}
\newcommand{\mRotateTrans}{\ensuremath{\vc{R}}}
\newcommand{\sJoint}{\ensuremath{q}}
\newcommand{\vJoint}{\ensuremath{\vc{q}}}
\newcommand{\mJoint}{\ensuremath{\vc{Q}}}
\newcommand{\mMass}{\ensuremath{\vc{M}}}
\newcommand{\sMass}{\ensuremath{{m}}}
\newcommand{\vGravity}{\ensuremath{\vc{g}}}
\newcommand{\vConstr}{\ensuremath{\vc{C}}}
\newcommand{\sConstr}{\ensuremath{C}}
\newcommand{\vCOM}{\ensuremath{\vc{x}}}
\newcommand{\sGeneralForce}[1]{\ensuremath{Q_{#1}}}
\newcommand{\vStateVar}{\ensuremath{\vc{y}}}
\newcommand{\vControlVar}{\ensuremath{\vc{u}}}
\newcommand{\argmax}{\operatornamewithlimits{argmax}}
\newcommand{\argmin}{\operatornamewithlimits{argmin}}

\newcommand{\tr}[1]{\ensuremath{\mathrm{tr}\left(#1\right)}}

%
%

\renewcommand{\choose}[2]{\ensuremath{\left(\begin{array}{c} #1 \\ #2 \end{array} \right )}}

\newcommand{\gauss}[3]{\ensuremath{\mathcal{N}(#1 | #2 ; #3)}}

\newcommand{\pctab}{\hspace{0.2in}}
\newenvironment{pseudocode} {\begin{center} \begin{minipage}{\textwidth}
                             \normalsize \vspace{-2\baselineskip} \begin{tabbing}
                             \pctab \= \pctab \= \pctab \= \pctab \=
                             \pctab \= \pctab \= \pctab \= \pctab \= \\}
                            {\end{tabbing} \vspace{-2\baselineskip}
                             \end{minipage} \end{center}}
\newenvironment{items}      {\begin{list}{$\bullet$}
                              {\setlength{\partopsep}{\parskip}
                                \setlength{\parsep}{\parskip}
                                \setlength{\topsep}{0pt}
                                \setlength{\itemsep}{0pt}
                                \settowidth{\labelwidth}{$\bullet$}
                                \setlength{\labelsep}{1ex}
                                \setlength{\leftmargin}{\labelwidth}
                                \addtolength{\leftmargin}{\labelsep}
                                }
                              }
                            {\end{list}}
\newcommand{\newfun}[3]{\noindent\vspace{0pt}\fbox{\begin{minipage}{3.3truein}\vspace{#1}~ {#3}~\vspace{12pt}\end{minipage}}\vspace{#2}}

\newcommand{\key}{\textbf}
\newcommand{\fun}{\textsc}



\section{Introduction}

Realistic movement of virtual humans plays an integral role in bringing fictional figures to life in films and games, enhancing immersive experiences in VR, and recently, teaching robots how to physically interact with humans \cite{Clegg:2018}. While such applications in robotics and machine learning have created new research avenues for the field of character animation, they also introduce new problems that challenge the existing techniques. Most noticeably, the virtual humans interacting with robots must exhibit not only human-like movements, but also valid joint torques consistent with their physiological capability.

\begin{figure}[t]
  \includegraphics[width=1.0\linewidth]{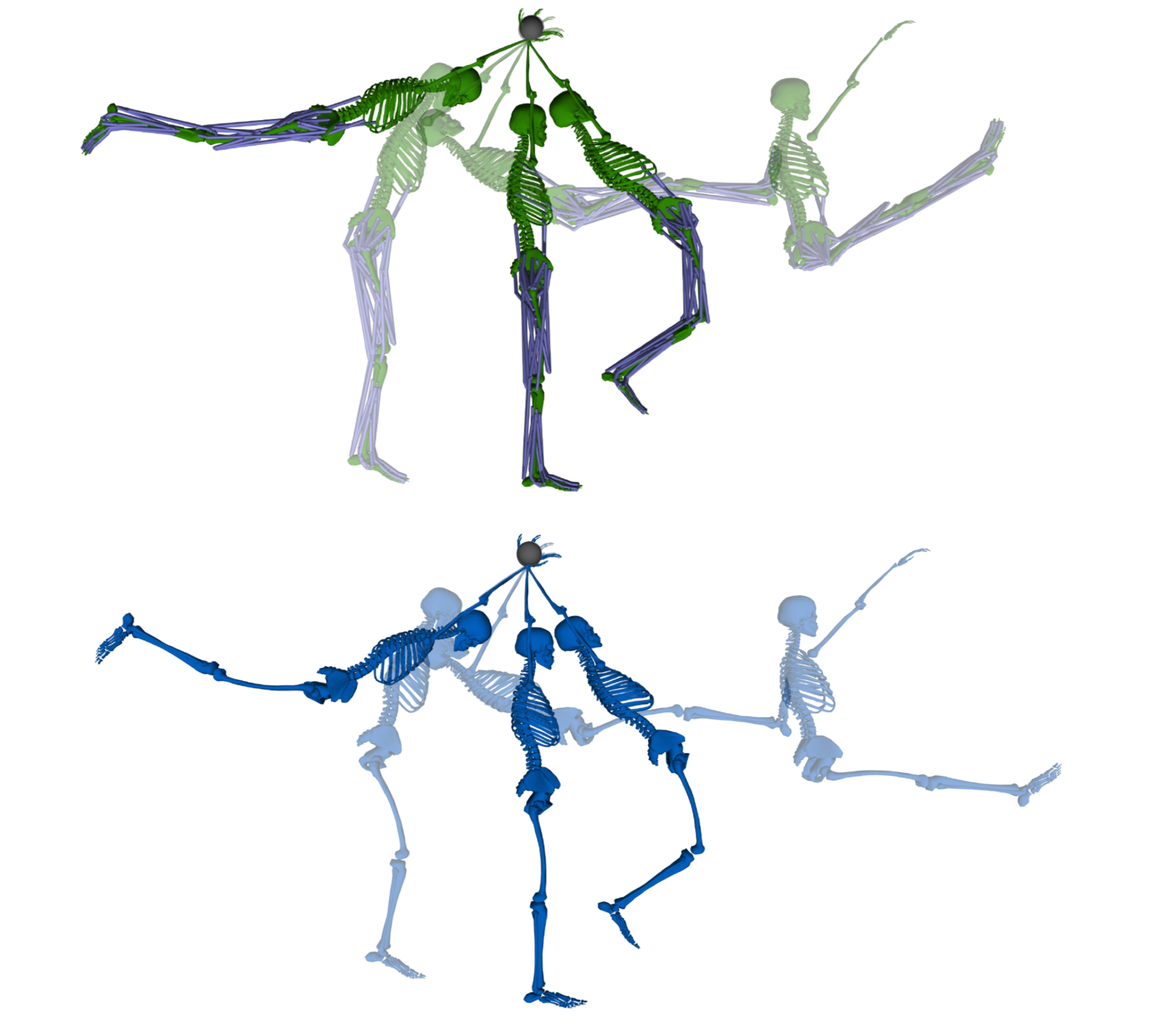}
  \caption{Top: A swing motion solved by trajectory optimization in the muscle-actuation space. Bottom: Our proposed method can solve the same task in the joint-actuation space, yielding similar motion but costing fewer iterations and less computation time.}
  \label{fig:swinging_max}
\end{figure}

Virtual human is often modeled as articulated rigid bodies with actuated joints that directly and independently generate torques to drive the kinematic movement. While \emph{joint-actuation} simplifies the modeling, simulation, and control of virtual humans, it often produces torque patterns that are unnatural or infeasible for real humans to achieve. Consequently, additional kinematic constraints or motion data are often needed to improve the naturalness of the kinematic trajectories. Alternatively, musculotendon models explicitly simulate the dynamics of muscles and tendons to drive the skeletal system. As such, models based on \emph{muscle-actuation} are able to impose physiologically realistic constraints and energetic cost on the resultant torque trajectories, leading to more human-like movements that reflect the mechanics of human anatomy. However, the realism of motion comes at the expense of complex modeling efforts and costly computation of simulation. As a result, large-scale trajectory optimization problems or sample-demanding reinforcement learning problems often eschew muscle-actuation and settle for simpler models that use joint-actuation.

\begin{figure}
  \includegraphics[width=1.0\linewidth]{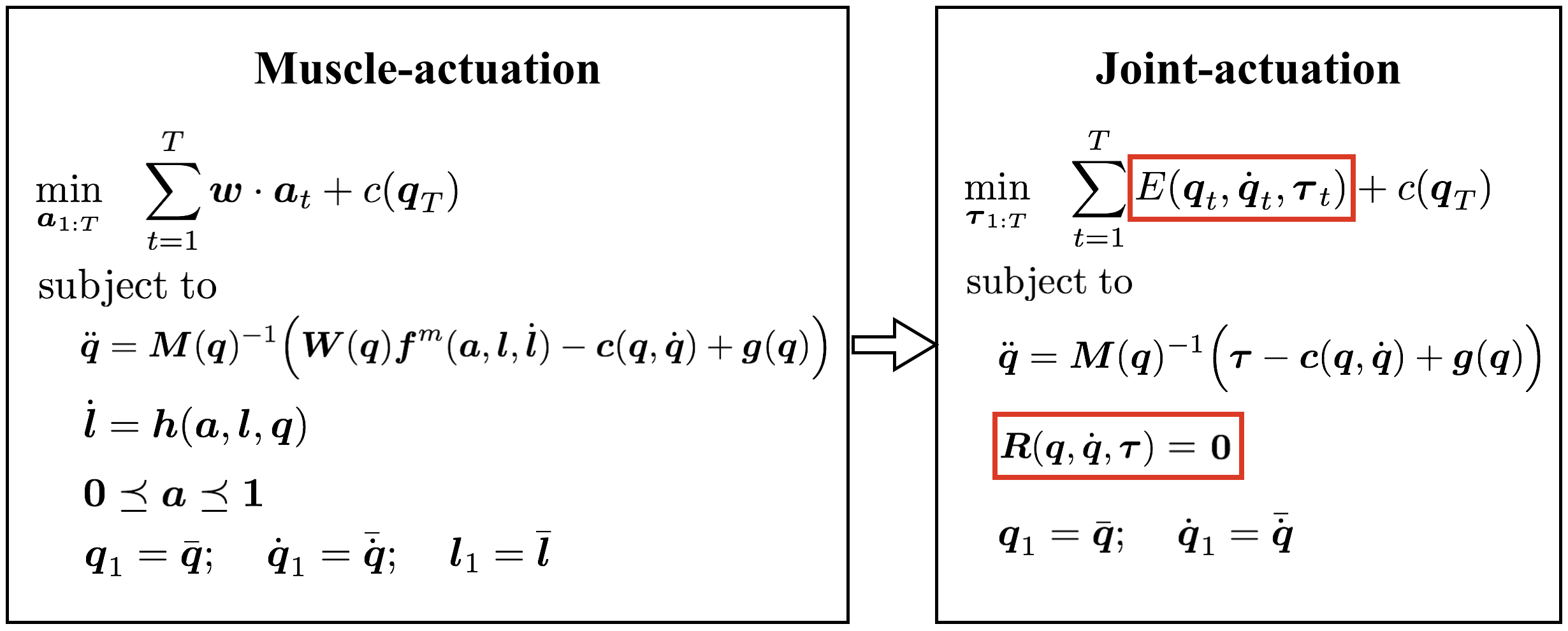}
  \caption{Left: A trajectory optimization problem formulated in the muscle-actuation space. The goal is to find a sequence of muscle activations $\vc{a}_{1:T}$ that minimizes the metabolic energy cost and other task-dependent performance terms (\eg the deviation from the target final state), subject to multibody dynamics, muscle contraction dynamics, muscle activation bounds, and the initial conditions. Right: A trajectory optimization problem formulated in the joint-actuation space. A sequence of joint torques $\boldsymbol{\tau}_{1:T}$ is solved directly under multibody dynamics. By learning joint torque limits $\vc{R}$ and the energy function $E$ from the muscle-based models, the two optimization problems can reach the same optimal value under some common assumptions.}
  \label{fig:two-optimal-problems}
\end{figure}

What if we can generate human-like motion comparable to muscle-actuation models, while retaining the benefit of simple modeling and fast computation offered by joint-actuation models? This paper introduces a technique to transform an optimal control problem formulated in the muscle-actuation space to an equivalent problem in the joint-actuation space, such that the solutions to both problems have the same optimal value. The class of optimal control problems addressed in this paper is general---it minimizes any objective function as long as it includes an energy penalty term involving muscle state and activation, subject to any set of constraints that includes equations of motion and bounds of muscle activations (Figure \ref{fig:two-optimal-problems} Left). If a motor control problem can be formulated in such a form, we can prove, under some common assumptions (detailed in Section \ref{sec:musculoskeletal-model}), that a torque trajectory with the same optimal value can be obtained by solving the equivalent problem formulated in the joint-actuation space, using much less computation time (Figure \ref{fig:two-optimal-problems} Right).

Comparing with an optimal control problem formulated in the muscle-actuation space, we identify two aspects in the standard joint-actuation formulation that lead to undesired torque solutions. First, torque limits are usually enforced by box constraints with empirically determined constant bounds. In reality, human joint torque limit is much more complex than a constant range; it depends on the position and velocity of the joint, as well as the position, velocity and actuated state of other joints. Second, the energy function typically penalizes the magnitude of torques, which does not account for the muscle anatomy and the effect of passive elements in the human musculoskeletal system. To bridge the gap, we propose a machine learning approach to approximate the state-dependent torque limits and the metabolic energy function parameterized in the joint-actuation space, using neural networks as function approximators (Figure \ref{fig:two-optimal-problems}, red boxes). The training data are generated via solving optimizations using a musculotendon model and a physiologically-based simulator, OpenSim \cite{seth:2011}. Once trained, the same torque limits and the energy function can be applied to drastically different motor tasks.


In addition to trajectory optimization, our technique can be applied to policy optimization using deep reinforcement learning (DRL) techniques. While the recent DRL techniques have demonstrated the possibility of learning human-like locomotion without motion data, the success of learning critically depends on the physical parameters of the skeleton, such as joint torque limits and the joint stiffness of the agent. The engineer has to tune the parameters for each degree of freedom for each task (\eg walking needs different joint torque limits from jumping), leading to an expensive trial-and-error process, while in reality these physical joint parameters of real humans are agnostic to the tasks. By replacing these task-specific, manually-determined parameters with our learned, state-dependent joint torque limits and energy function, we provide more anatomically realistic action space and energy function to guide the agent's explorations during its learning of motor tasks.

To show that our technique can be applied to both trajectory optimization and policy learning, we solve collocation problems for jumping and swinging, and learn policies using policy gradient methods for walking and running. We also evaluate the advantages of our technique over conventional box torque limits and sum-of-torque energy penalty, as well as comparing to motions generated by muscle-based models. 

\section{Related work}
\label{sec:related-work}
 In computer animation, trajectory optimization, or spacetime constraints \cite{Witkin:1988}, provides a general mathematical tool for motion synthesis \cite{Rose:1996,Popovic:1999,Liu:2002,Fang:2003,Safonova:2004,Mordatch:2012}, character retargeting \cite{Gleicher:1998,Wampler:2014}, or style editing \cite{Liu:2005,Kass:2008,Min:2010}. Using the same optimization framework, this paper shows that by replacing the conventional energy term and the box torque limits, our method can reduce the need for parameter tuning and improve the quality of the motion. Our method can also be utilized by any model-free reinforcement learning (RL) framework. Early researchers in the graphics community exploited RL techniques to learn high-level decision making for motor skills \cite{Lee:2004,Treuille:2007,McCann:2007,Lo:2008,YLee:2009,Coros:2009}. Recently, researchers leveraged the representative power of neural networks to train policies that directly map high-dimensional sensory input to actuation for complex motor tasks \cite{Peng:2017,Peng:2018,Won:2018,Clegg:2018}. Similar to trajectory optimization, our learned energy function can replace the common energy or regularization term in the reward function. Our learned torque limits are agnostic to the learning algorithm as it only impacts the transition dynamics.
 
Researchers in biomechanics have developed musculoskeletal models that use biomimetic muscles and tendons to drive skeletal motion. In contrast to joint-actuation models, muscle-actuation results in movements that respect physiological constraints \cite{Komura:2000} and exhibit energy expenditure more similar to real humans \cite{Wang:2012}. In addition, simulating the behavior of passive elements has been shown to improve the stability of the movements against perturbations \cite{Gerritsen:1998,VanDerKrogt:2009}. Controlling a muscle-based virtual character has also been explored in computer animation. In the past two decades, researchers have demonstrated increasingly more impressive results from upper body movements \cite{Lee:2006,Lee:2009,Lee:2018}, to hand manipulation \cite{Tsang:2005,Sueda:2008}, to facial animation \cite{Lee:1995,Sifakis:2005}, and to locomotion \cite{Wang:2012,Geijtenbeek:2013,Mordatch:2013,Lee:2014,Si:2014}. Recently, Nakada \etal \shortcite{Nakada:2018} introduced a sensorimotor system that directly maps the photoreceptor responses to muscle activations, bestowing a visuomotor controller of the character's eyes, head, and limbs for non-trivial, visually-guided tasks. Of particular interest to us is the use of training data synthesized by a muscle-based simulator \cite{Lee:2006}. Although the focus of our work is orthogonal to learning control policies with a muscle-based simulator, we also take advantage of physiologically-based simulators for generating training data that would otherwise be infeasible to acquire in the real world. When transforming a problem from the muscle-actuation to the joint-actuation space, we solve muscle redundancy problems \cite{de2016evaluation}, assuming human joint torque is produced by muscles with certain performance criteria optimized \cite{pedotti1978optimization}. Peng and van de Panne \shortcite{peng2017learning} compared how joint-actuation or muscle-actuation control affects the performance of RL in tracking reference motions.

One important advantage of the muscle-based model is that it provides physiologically realistic joint torque limits based on muscle physiology and the anatomy of human musculoskeletal system where muscles often span multiple joints. In contrast, joint-actuation models rely on the engineers to manually set the torque limits independently for each joint, resulting in torque patterns infeasible for humans to achieve \cite{Komura:2000,Geijtenbeek:2010}. Recent work by Yu \etal \shortcite{Yu:2018} also reported that DRL methods result in policies sensitive to the range of action space (\ie the joint torque limits), which is often arbitrarily or empirically determined. In reality, the torque limits observed at each joint are due to the complex interplay of multiple muscle-tendon units actuating a single or multiple joints \cite{delp:1990}. Studies in biomechanics and ergonomics showed that the ranges of torques at each joint depend on the positions and velocities of the joint, as well as those of other joints \cite{Amis:1980,Nijhofa:2006,anderson:2007}. For example, the maximum force the shoulder can generate depends on the hand position relative to the shoulder. 

The energy function or performance criterion plays a crucial role in the optimal control of human movements. In computer animation, a common practice is to use the sum of squared joint torques. The weighting of the joints can be determined based on the inertial properties of the character \cite{Popovic:1999} or the task performed by the character \cite{Liu:2005,Ye:2008}.  The equivalent of minimizing squared torques for muscle-driven simulation would be minimizing squared muscle forces. However, minimization of squared muscle forces does not result in realistic muscle coordination patterns because this criterion does not account for the dependence of muscle's force generating capacities on its length and velocity \cite{pedotti1978optimization}. In contrast, the above-mentioned issues can be mitigated by musculotendon-based energy function which approximates the metabolic energy expenditure \cite{umberger2003model, umberger2010stance, bhargava2004phenomenological}. More human-like torque patterns were also observed when metabolic energy expenditure is minimized \cite{Wang:2012}. Simulated human walking patterns in three dimensions are sensitive to the choice of a muscle energy model \cite{miller2014comparison}. Another popular performance criterion in biomechanics is minimizing sum of muscle activations to a power of 1 to infinity. This criterion also favors muscles that operate close to their optimal lengths and velocities. Fatigue-like cost functions that minimize peak muscle activity (higher powers) predicted more realistic gait patterns in 2D simulations than energy-like cost functions that minimize muscle activation squared \cite{ackermann2010optimality}. 

The Learning to Run Challenge applied DRL approaches to learning a running policy on a muscle-based model \cite{DBLP:journals/corr/abs-1804-00198,Kidzinski:2018}. The task for the participants was to develop a successful policy to enable a physiologically-based human model moving as fast as possible on an obstacle course. Although the human model is simplified to 2D with only $9$ degrees-of-freedom actuated by $18$ muscle actuators, with modifications and improvements to the off-the-shelf RL algorithms, many participants succeeded in showing robust and efficient running gaits. For the leading teams who reported their computation time in \cite{Kidzinski:2018}, most policies took a few days to train and the computation bottleneck is mainly on the muscle-based simulation. Our work poses an even more challenging learning problem. The human musculoskeletal model we use has 23 degrees-of-freedom, actuated by 92 muscle-tendon actuators, and it is fully in 3D. Directly using state-of-the-art policy learning algorithms on such a model is not feasible.
\section{Method}

We propose a method to train two functions represented as neural networks to model the state-dependent torque limits and the metabolic energy function. The trained function approximators are able to faithfully reflect biological properties achieved by explicit muscle modeling, but the input of the functions only involves quantities available in the joint-actuation space. We utilize a muscle-based human model and the OpenSim simulator to generate training data for learning the neural networks. Once trained, the two task-agnostic, analytical and differentiable functions can be used to solve general optimal control problems of human motion in the joint actuation space.

\subsection{Human Musculoskeletal Model}
\label{sec:musculoskeletal-model}

\begin{figure}
  \includegraphics[width=1.0\linewidth]{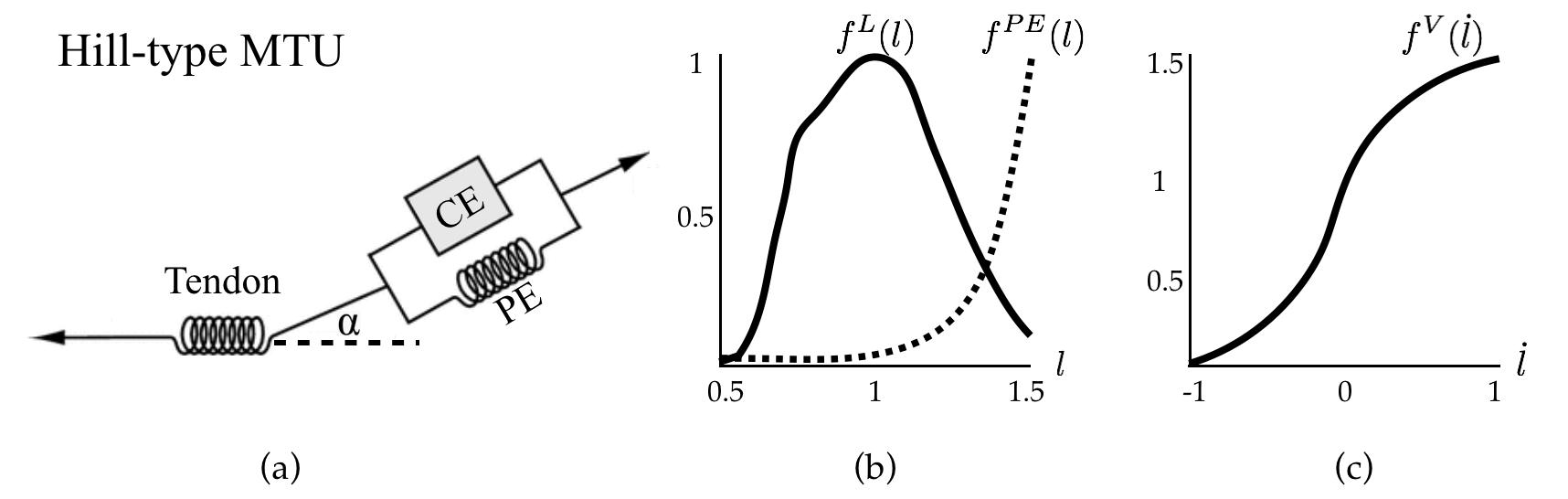}
  \caption{(a) Hill-type model illustration. We assume that the change of tendon length is negligible and model the tendon as an inelastic wire. (b) Muscle force-length curve and passive muscle force (dashed curve). (c) Muscle force-velocity curve.} 
  \label{fig:muscle-model}
\end{figure}

The human figure is represented as a musculoskeletal model using \emph{gait2392} \cite{delp:1990} provided by OpenSim. This model consists of a skeleton of legs and a torso without arms and a head, connected by joints that amount to $23$ degrees-of-freedom (DOFs). $92$ muscle-tendon actuators (MTUs) are used to represent $76$ muscles in the lower extremities and the torso \cite{delp:1990}. Through the predefined muscle path, each MTU is attached to two or more bone segments. It is important to note that each DOF is driven by multiple MTUs and each MTU can actuate multiple DOFs spanning multiple joints.

The force generated by each MTU is computed based on a Hill-type model \cite{zajac:1989}, composing an active contractile element (CE) in parallel connection with a passive element (PE), and a passive tendon element in series with the CE and PE (Figure \ref{fig:muscle-model}(a)). The CE generates active contraction force with the magnitude proportional to the level of activation $a \in [0,1]$, and the PE generates a passive, non-linear spring force. 

To transform an optimal control problem from the muscle-actuation space to the joint-actuation space, we make two assumptions about the Hill-type MTUs, both of which are commonly acceptable in biomechanics and computer animation communities \cite{Lee:2009,Lee:2014,anderson2001static,Lee:2018}. First, we do not model tendon compliance, which implies that a tendon is modeled as an inelastic wire with negligible length change. Second, we do not consider the activation dynamics and use activation instead of neural excitation as the control input. The implications of both simplifications will be further discussed in Section \ref{sec:discussion}. 

With the simplifications, the muscle force generated by each MTU is given by the sum of CE force and PE force:
\begin{equation}
    f^{m} = f_o^{m} \Big(a f^L(l) f^V(\dot{l}) + f^{PE}(l)\Big) \cos \alpha,
    \label{eqn:muscle-force}
\end{equation}
where $l$, $\dot{l}$ are normalized muscle fiber length and velocity, and $\alpha$ is the muscle pennation angle, which value depends on muscle length. The constant $f_o^{m}$ is the optimal muscle fiber force, generally proportional to the volume of muscle. Figure \ref{fig:muscle-model} shows the functions $f^L$, $f^V$ and $f^{PE}$ based on in-situ experiments \cite{gollapudi:2009, joyce:1969}. All the constants and normalization factors in Equation \ref{eqn:muscle-force} can be found in biomechanics literature \cite{delp:1990, anderson:2001}. Once muscle forces are obtained, we compute the joint torque for each DOF by summing up the contribution of every muscle spanning that DOF: 
\begin{equation}
\label{eq:muscle-dynamics}
    \tau_i = \sum_{j=1}^{n_m} \vc{W}(\vc{q})[i,j] f^{m}_j,
\end{equation}
where $\vc{q}$ is the joint configuration of the skeleton and $\vc{W}(\vc{q}) \in \mathbb{R}^{n_q \times n_m}$ is the moment arm matrix that transforms muscles forces into the joint coordinates. We denote the number of muscles and the number of DOFs as $n_m$ and $n_q$ respectively. If the $j$-th muscle does not actuate the $i$-th DOF, the corresponding element, $\vc{W}(\vc{q})[i,j]$, will be zero.

Since we assume that the tendon length is fixed, the length $l_j$ and velocity $\dot{l}_j$  of each muscle can be fully determined by the joint angles $\vc{q}$ and velocities $\dot{\vc{q}}$. We then define the muscle dynamics function (Equation \ref{eq:muscle-dynamics}) as $\boldsymbol{\tau} = \vc{D}(\vc{a}, \vc{q}, \dot{\vc{q}})$, where $\vc{a} \in \mathbb{R}^{n_m}$ and $\bm{\tau}, \vc{q}, \dot{\vc{q}} \in \mathbb{R}^{n_q}$. The torques generated by the MTUs will then be integrated along with other forces according to the equations of motion for the articulated skeleton:
\begin{equation}
\vc{M}(\vc{q})\ddot{\vc{q}} + \vc{c}(\vc{q}, \dot{\vc{q}}) = \boldsymbol{\tau} + \vc{g}(\vc{q}) + \vc{J}(\vc{q})^T\vc{f},    
\end{equation}
where $\vc{M}$ is the mass matrix, $\vc{c}$ is the Coriolis force, $\vc{g}$ is the gravitational force, $\vc{f}$ indicates other forces applied to the system, and $\vc{J}$ is the Jacobian matrix that transforms the forces to the joint coordinates. As a short-hand, we define the skeletal dynamic function as $\ddot{\vc{q}} = \vc{S}(\boldsymbol{\tau}, \vc{q}, \dot{\vc{q}})$. Note that $\vc{S}$ does not include other forces $\vc{f}$.


\subsection{Learning state-dependent torque limits}

Consider the inverse problem of muscle dynamics---given a torque vector $\boldsymbol{\tau}$, can we determine whether $\boldsymbol{\tau}$ is realizable by the human muscles? To answer this question, we need to consider not only the current joint state ($\vc{q}, \dot{\vc{q}}$), but also the inter-dependency of joint torques due to muscle sharing. As such, we define the state-dependent torque limits as an implicit equation:
\begin{equation}
    C(\vc{q}, \dot{\vc{q}}, \boldsymbol{\tau}) = \begin{cases} -1, & \mathrm{~~if~~} \exists~ \vc{0} \preceq\vc{a} \preceq \vc{1}, \mathrm{\;\;\;s.t.\;} \boldsymbol{\tau}= \vc{D}(\vc{a}, \vc{q}, \dot{\vc{q}}), \\ 1, & \mathrm{~~otherwise}, \end{cases}
\end{equation}
where the input $\boldsymbol{\tau}$ is the proposed torque to be validated. The feasibility function $C$ returns $-1$ if the proposed torque is human-feasible, and $1$ if not. We could then train a differentiable function to approximate $C$ and use it as a constraint in a trajectory optimization problem. However, in the reinforcement learning setting, enforcing a hard constraint $C(\vc{q}, \dot{\vc{q}}, \boldsymbol{\tau}) \leq 0$ during rollout simulation can be challenging and inefficient. 

To mitigate the above issue, we instead build an equivalent \emph{correction} function $\vc{R}$:
\begin{equation}
    \label{eq:tor-limit-optimize}
    \vc{R}(\vc{q}, \dot{\vc{q}}, \boldsymbol{\tau}) = \argmin_{\Delta \boldsymbol{\tau}} \norm{\Delta \boldsymbol{\tau}}_2 \mathrm{\;\;\;s.t.\;} \exists~ \vc{0} \preceq\vc{a} \preceq \vc{1},~~ \boldsymbol{\tau} - \Delta \boldsymbol{\tau}= \vc{D}(\vc{a}, \vc{q}, \dot{\vc{q}}),
\end{equation}
where $\Delta \boldsymbol{\tau}$ is the difference between an infeasible torque vector $\boldsymbol{\tau}$ and its L2-closest feasible neighbor. A torque vector $\boldsymbol{\tau}$ is feasible if $\vc{R} = \vc{0}$, otherwise \vc{R} outputs the correction vector $\Delta \boldsymbol{\tau}$. For trajectory optimization, instead of applying the constraint $\vc{R} = \vc{0}$, a slightly relaxed version $-\boldsymbol{\epsilon} \preceq \vc{R} \preceq \boldsymbol{\epsilon}$ can be used in practice to account for the regression errors of the function approximator. For reinforcement learning, we can now effectively enforce the validity of torques in the physics simulator by projecting an invalid torque proposed by the policy to a valid one on the boundary of the torque limits: $\boldsymbol{\tau} - \vc{R}(\vc{q}, \dot{\vc{q}}, \boldsymbol{\tau})$. 


Using the OpenSim lower-extremity muscle model \emph{gait2392}, and assuming the torque limits of one leg is independent of the other, $\vc{R}$ maps from $\mathbb{R}^{15}$ to $\mathbb{R}^5$, as $\vc{q}$, $\dot{\vc{q}}$ and $\boldsymbol{\tau}$ are all in $\mathbb{R}^5$ \footnote{For each leg, we include three DOFs for hip, one DOF for knee, and one DOF for ankle.}. We parameterize $\vc{R}$ as a feed-forward neural network with three hidden layers with 180-128-64 neurons. ELU activation \cite{clevert:2015} is used for the hidden layers to ensure differentiability. The input of training data is generated by uniformly sampling $1M$ vectors of $(\vc{q}, \dot{\vc{q}}, \boldsymbol{\tau})$ in $\mathbb{R}^{15}$, with bounded range in each dimension. The range of each dimension is approximately determined referring to collected motion data and biomechanics literature \cite{pandy1990optimal,anderson1999dynamic} such that it covers the human joint limits, joint velocity limits, or torque limits. For example, $\vc{q}_1$ (hip flexion angle) is uniformly sampled (1M times) within $-0.9$ rad and $2$ rad; $\dot{\vc{q}}_4$ (knee velocity) is sampled within $-10$ rad/s and $10$ rad/s; and $\bm{\tau}_4$ (knee torque) is sampled within $-250$ Nm and $250$ Nm. The output of each training point is generated by solving right-hand side of Equation \ref{eq:tor-limit-optimize} using an optimization solver IPOPT \cite{wachter:2006}. The neural network training takes $500$ epochs and is able to reach $95\%$ accuracy on a $150K$ independently sampled test set.


Once trained, we can use the neural network to approximate the output of the costly optimization of Equation \ref{eq:tor-limit-optimize} and compute the gradient of $\vc{R}$.
The forward pass and back-propagation of the trained neural networks add negligible overhead to each iteration of control optimization.

\subsection{Learning muscle-based energy rate function}

An important advantage of muscle-actuation models is that physiologically based energy function can be easily formulated using muscle activations and the muscle states:
\begin{equation}
    \textrm{Total~Effort} = \int_0^T \sum_{j=1}^{n_m} p_j(a_j, l_j, \dot{l}_j) ~\mathrm{d}t,
\end{equation}
where $p_j$ is the energy rate function for muscle $j$. In general, different muscles have the same form of $p_j$ except for individual scaling factors. Various formulae of $p_j$ have been proposed (Section \ref{sec:related-work}) but consensus has not been reached in the biomechanics community. However, muscle-based energy rate formulae are expected to be more accurate than the sum-of-torques formula commonly used in the joint-actuation formulation, because they can easily exclude forces generated by passive MTU elements from the energy calculation.

Is it possible to recover muscle-based energy rate given only quantities available in the joint-actuation space, namely, $\vc{q}$, $\dot{\vc{q}}$ and $\boldsymbol{\tau}$? With our rigid tendon assumption, the muscle state ($\vc{l}, \dot{\vc{l}}$) can be derived from ($\vc{q}, \dot{\vc{q}}$). However, at a certain state $(\vc{q}, \dot{\vc{q}})$ with an applied human-feasible torque $\boldsymbol{\tau}$, there exists infinite combinations of muscle activations that can realize $\boldsymbol{\tau}$ since the muscle system is over-actuated. To resolve the redundancy, we assume that human generates the minimal-energy muscle activation pattern for a given $(\vc{q}, \dot{\vc{q}}, \boldsymbol{\tau})$:
\begin{equation}
    E(\vc{q}, \dot{\vc{q}}, \boldsymbol{\tau}) = \min_{\vc{0} \preceq\vc{a} \preceq \vc{1}} \sum_{j=1}^{n_m} \hat{p}_j(a_j, \vc{q}, \dot{\vc{q}})  \mathrm{\;\;\;s.t.\;} \boldsymbol{\tau}= \vc{D}(\vc{a}, \vc{q}, \dot{\vc{q}}),
    \label{eqn:energy-function}
\end{equation}
where $\hat{p}_j$ is energy rate function parameterized by joint state and muscle activation. We choose a simple energy rate formula:
\begin{equation}
\sum_{j=1}^{n_m} \hat{p}_j(a_j, \vc{q}, \dot{\vc{q}}) = \sum_{j=1}^{n_m} f_{o_j}^m \cdot a_j = \vc{w} \cdot \vc{a},
\end{equation}
where the scaling factors $\vc{w} = (f_{o_1}^m, \cdots f_{o_{n_m}}^m)$ (ref. Equation \ref{eqn:muscle-force}) make larger muscles more costly to generate forces. 

To train a neural network to approximate $E$, we need to sample the space of $(\vc{q}, \dot{\vc{q}}, \boldsymbol{\tau})$ and compute the corresponding energy value by solving the right-hand side of Equation \ref{eqn:energy-function}. However, we cannot naively sample the space of $(\vc{q}, \dot{\vc{q}}, \boldsymbol{\tau})$ as $E$ is only defined for the feasible torques. Instead, we uniformly sample $1.5$M vectors of $(\vc{q}, \dot{\vc{q}}, \vc{a})$, within $\vc{0} \preceq\vc{a} \preceq \vc{1}$ and reasonably large bounds for $\vc{q}$ and $\dot{\vc{q}}$. Through the muscle dynamics function, $\boldsymbol{\tau} = \vc{D}(\vc{a}, \vc{q}, \dot{\vc{q}})$, we can recover $1.5$M feasible torques $\boldsymbol{\tau}$. Concatenating $\boldsymbol{\tau}$ with corresponding $(\vc{q}, \dot{\vc{q}})$, we then solve Equation \ref{eqn:energy-function}  using IPOPT to obtain the output of training data. Note that the uniformly sampled $\vc{a}$'s are discarded as they are likely not the most efficient activation patterns to generate the corresponding $\boldsymbol{\tau}$'s. 

We represent $E$ as a feed-forward neural network that maps $\mathbb{R}^{15}$ to $\mathbb{R}$ with three hidden layers of 360-180-80 neurons. The training takes $500$ epochs and is able to reach $92\%$ accuracy on a $150$K independently sampled test set. Once trained, the computation required for solving Equation \ref{eqn:energy-function} can be reduced to one forward pass of the neural network. The gradient of $E$ can be obtained through back-propagation.

\subsection{Proof of equivalent optimal value}
To prove that the two trajectory optimization problems in Figure \ref{fig:two-optimal-problems} lead to solutions with the same optimal value, we first redefine the optimal control problem in the muscle-actuation space with the rigid tendon assumption: 
\begin{eqnarray*}
\min_{\mathcal{A}} & L_a(\mathcal{A}) = \sum_{t=1}^T \vc{w} \cdot \vc{a}_t + c(\vc{q}_T), \\ 
    \mathrm{subject\;to} & \ddot{\vc{q}}_t = \vc{S}(\boldsymbol{\tau}_t, \vc{q}_t, \dot{\vc{q}}_t),  \\
    & \boldsymbol{\tau}_t= \vc{D}(\vc{a}_t, \vc{q}_t, \dot{\vc{q}}_t),  \\
    & \vc{0} \preceq\vc{a}_t \preceq \vc{1}, \\
    &  \vc{q}_1 = \bar{\vc{q}};\; \dot{\vc{q}}_1 = \bar{\dot{\vc{q}}},
\end{eqnarray*}
where $\mathcal{A} := \{\vc{a}_1, \cdots, \vc{a}_T\}$, $(\bar{\vc{q}}, \bar{\dot{\vc{q}}})$ is the given initial state, and without loss of generality, we assume that the objective function is composed of an effort term that minimizes the weighted sum of muscle activation and a task term that depends on the final state $\vc{q}_T$. For clarity, we omit the range of subscript $t$ for each constraint and assume it to be $1 \leq t \leq T$ unless otherwise specified. We also omit implicit decision variables $\vc{q}$ and $\dot{\vc{q}}$, as they depend on the control variables $\vc{a}$, given the initial boundary conditions and the dynamical constraints. The proposed equivalent problem for optimizing the torque trajectory $\mathcal{T} := \{\boldsymbol{\tau}_1, \cdots, \boldsymbol{\tau}_T \}$ in the joint-actuation space is then denoted as:
\begin{eqnarray*}
\min_{\mathcal{T}} & L_\tau(\mathcal{T}) = \sum_{t=1}^T E(\vc{q}_t, \dot{\vc{q}}_t, \boldsymbol{\tau}_t) + c(\vc{q}_T), \\
    \mathrm{subject\;to} & \ddot{\vc{q}}_t = \vc{S}(\boldsymbol{\tau}_t, \vc{q}_t, \dot{\vc{q}}_t), \\
    & \vc{R}(\vc{q}_t, \dot{\vc{q}}_t, \boldsymbol{\tau}_t) = \vc{0}, \\
    & \vc{q}_1 = \bar{\vc{q}};\; \dot{\vc{q}}_1 = \bar{\dot{\vc{q}}}.
\end{eqnarray*}

Our goal is to show that $\min_{\mathcal{T}} L_\tau(\mathcal{T}) = \min_{\mathcal{A}}  L_a(\mathcal{A})$ at the global minimum. Let us consider any minimizer of $L_\tau$\footnote{There could be multiple distinct minimizers giving the same optimal value.}, $\mathcal{T}^* = \argmin_{\mathcal{T}} L_\tau(\mathcal{T})$, and the state trajectory $(\vc{q}^{\tau^*}_{1:T}, \dot{\vc{q}}^{\tau^*}_{1:T})$ it generates. For each $\tau_t^*$ in $\mathcal{T}^*$, we can compute its minimal-energy muscle activation as:
\begin{equation}
    \vc{a}^{\tau^*}_t = \argmin_{\vc{0} \preceq\vc{a} \preceq \vc{1}} \vc{w} \cdot \vc{a} \mathrm{\;\;\;s.t.\;} \boldsymbol{\tau}_t^* = \vc{D}(\vc{a}, \vc{q}^{\tau^*}_t, \dot{\vc{q}}^{\tau^*}_t),
\end{equation}
and denote $\mathcal{A}^{\mathcal{T}^*} := \{\vc{a}^{\tau^*}_1, \cdots, \vc{a}^{\tau^*}_T\}$. Since both $\mathcal{A}^{\mathcal{T}^*}$ and $\mathcal{T}^*$ produce the same state trajectory, and $E$ returns the energy of the most energy-efficient muscle activation for a given $(\vc{q}, \dot{\vc{q}}, \boldsymbol{\tau})$, it is trivial to conclude that
\begin{equation}
L_a(\mathcal{A}^{\mathcal{T}^*}) = L_\tau(\mathcal{T}^*).
\label{eqn:induced-muscle}
\end{equation} 

Now let us consider any minimizer of $L_a$, $\mathcal{A}^* = \argmin_\mathcal{A} L_a(\mathcal{A})$, and denote $\mathcal{T}^{\mathcal{A}^*}$ to be the torque sequence generated by forward simulating $\mathcal{A}^*$ from $(\bar{\vc{q}}, \bar{\dot{\vc{q}}})$. Since both $\mathcal{A}^*$ and $\mathcal{T}^{\mathcal{A}^*}$ produce the same state trajectory $(\vc{q}^{a^*}_{1:T}, \dot{\vc{q}}^{a^*}_{1:T})$, and  $\vc{w} \cdot \vc{a}^*_t \geq E(\vc{q}^{a^*}_t, \dot{\vc{q}}^{a^*}_t, \boldsymbol{\tau}^{a^*}_t) $ for each $\vc{a}^*_t \in \mathcal{A}^*$ and $\boldsymbol{\tau}^{a^*}_t \in \mathcal{T}^{\mathcal{A}^*}$, we conclude that
\begin{equation}
    L_a(\mathcal{A}^*) \geq L_\tau(\mathcal{T}^{\mathcal{A}^*}).
    \label{eqn:induced-torque}
\end{equation}

Then, since $\mathcal{A}^*$ is the minimizer of $L_a$ and $\mathcal{T}^*$ is the minimizer of $L_\tau$, together with Equation \ref{eqn:induced-torque}, we have the following relations:
\begin{equation}
    L_a(\mathcal{A}^{\mathcal{T}^*}) \geq L_a(\mathcal{A}^*) \geq L_\tau(\mathcal{T}^{\mathcal{A}^*}) \geq L_\tau(\mathcal{T}^*).
    \label{eqn:four-relations}
\end{equation}

Considering Equation \ref{eqn:induced-muscle}, all four terms in Equation \ref{eqn:four-relations} must be equal. Therefore, we arrive at $\min_{\mathcal{T}} L_\tau(\mathcal{T}) = \min_{\mathcal{A}}  L_a(\mathcal{A})$. $\;\;\;\;\;\;\;\Box$


\subsection{Implementation}
The two neural networks, $\vc{R}$ and $E$ only need to be trained once for a particular human model and can be used in trajectory optimization or reinforcement learning for various motor tasks. We describe the implementation details below.

\subsubsection{Trajectory optimization}
The learned torque limits $\vc{R}$ can be readily used as constraints in optimal control problems to ensure the feasibility of torque trajectory. In our implementations, we allow a threshold of $\vc{-3} \preceq\vc{R} \preceq \vc{3}\textrm{~(Nm)}$ to account for the error due to function approximation.

Applying the learned energy function $E$ in trajectory optimization is more involved. Many optimization methods, such as interior-point method, allow constraints to be violated at early iterations of optimization, which can lead to infeasible $\vc{\tau}$. Since $E$ is only defined when $\vc{\tau}$ is feasible (\ie $C(\vc{q}, \dot{\vc{q}}, \boldsymbol{\tau}) \leq 0$), we cannot expect $E$ to return an accurate energy expenditure during early iterations. To mitigate this issue, we define an augmented function of $E$ for both feasible and infeasible regions:
\begin{equation}
    \tilde{E}(\vc{q}, \dot{\vc{q}}, \boldsymbol{\tau}) =  
    \begin{cases} 
    E(\vc{q}, \dot{\vc{q}}, \boldsymbol{\tau}), \;\;\;\; \mathrm{~~if~~} C(\vc{q}, \dot{\vc{q}}, \boldsymbol{\tau})\leq 0, \\ 
    E(\vc{q}, \dot{\vc{q}}, \boldsymbol{\tau}-\vc{R}(\vc{q}, \dot{\vc{q}}, \boldsymbol{\tau}))+ w \norm{\vc{R}(\vc{q}, \dot{\vc{q}}, \boldsymbol{\tau})}_2, 
    \;\;\;\mathrm{~~otherwise}, 
    \end{cases}
\end{equation}
where $w$ is set to a small constant discouraging use of infeasible torques. We train a neural network to approximate $\tilde{E}$ and use it in the objective function. During trajectory optimization, each iteration requires computing the gradients of objective function and the constraint Jacobian. Since our neural networks $\vc{R}$ and $\tilde{E}$ are both small, the computation overhead to the evaluation routines is negligible. 


\subsubsection{Reinforcement learning}
\label{sec:RL-implement}
The policy learning problem can be formulated as a Markov Decision Process and solved by model-free RL approach. The goal of this approach is to learn a policy that maps a state to an action, which can be a muscle activation vector if a muscle-based model is used, or a torque vector if a simpler joint-actuation model is used. In both cases, the physics simulator for the model then steps forward with the action computed by the current policy. 

Our method is agnostic to specific policy learning algorithm because it only modifies the physics simulation and the reward calculation. For each simulation step, since the torque $\boldsymbol{\tau}_p$ commanded by the policy could be infeasible, we clip $\boldsymbol{\tau}_p$ to $\boldsymbol{\tau}_p - \vc{R}(\vc{q}, \dot{\vc{q}}, \boldsymbol{\tau}_p)$ using our trained $\vc{R}$. A reward is calculated after each simulation step, where we use $E$ as an effort term in the reward function. Note that we do not need the augmented function $\tilde{E}$ because we use the clipped torque as the input to $E$ and thus is always in the feasible region of $E$. We also find that adding $-w \norm{\vc{R}}_2$ ($w>0$) to the reward function to penalize the use of correction torque helps the learning. We keep $w$ a fixed constant in our experiments. Gradient computation of $\vc{R}$ and $E$ is not needed in model-free policy learning.
\section{Evaluation}
We evaluate our learned torque limits and the energy function on trajectory optimization and policy learning problems using four distinct motor skills---jumping and swinging for demonstrating trajectory optimization, and walking and running for demonstrating policy learning. The experiments below aim to validate the following advantages of our method. First, with our method, the optimal control problems formulated in the joint-actuation space is able to produce comparably natural motions to those explicitly solved by optimizing muscle activation with complex musculotendon models, but using less computation time. Second, comparing to the commonly used box torque limits, our method produces more human-like movements and torque patterns, and as well eliminates the need to tune the torque limits for each specific task. Third, our method lends itself well to policy learning with DRL. Comparing to musculotendon models, the joint-actuation formulation reduces the dimension of the action space, making deep reinforcement learning more tractable.

We conduct ablation study on our two main components, the learned torque limits and the learned energy function. LR+LE refers to the evaluation using both $\vc{R}$ and $E$, while LR only uses the torque limits $\vc{R}$. In the case of LR, the effort cost, if needed by the control problem, is calculated by summing the absolute value of torques. We also develop three baselines to compare our method against: 
\begin{enumerate}
    \item{BOX:} Use joint-actuation with a constant range of torque for each DOF.
    \item{MTU:} Use $86$ MTUs to actuate the  lower-limb DOFs, subject to muscle activation limits, $\vc{0} \preceq \vc{a} \preceq \vc{1}$.
    \item{AMTU:} A method to accelerate MTU by training a regressor represented as a neural network to approximate the forward muscle dynamics function $\boldsymbol{\tau} = \vc{D}(\vc{a},\vc{q},\dot{\vc{q}})$. AMTU is used only for reinforcement learning. More details in Section \ref{sec:policy-learning}.
\end{enumerate}

\subsection{Trajectory optimization}
We use a 2D human model with three DOFs that flex/extend the hip, knee, and ankle to solve the trajectory optimization problems. For the BOX baseline, we set the torque limit of all three DOFs to $[-200,200]$ Nm, which lies within human capability \cite{anderson1999dynamic}. The trajectory optimization problems are formulated as a direct collocation \cite{rao:2009} solved by IPOPT with our implementation in Matlab 2017b \cite{Matlab:2017}.

For both jumping and swinging problems, the initial guess for the state variables is simply a constant trajectory holding the pose of the first frame. The initial guess for the control variables is also a constant trajectory with a small torque for each DOF. Note that the initial state and control trajectories do not need to be dynamically consistent. For all our experiments, the optimization is terminated when the constraint violation is sufficiently small and the objective value varies less than $0.1\%$ for $10$ consecutive iterations. 

\subsubsection{Jumping}
The 2D jumper is modeled as a four-link rigid-body chain with foot, shin, thigh, and torso, whose goal is to jump as high as possible from a pre-specified crouching pose with zero initial velocity. We divide the vertical jump to two phases: ground-contact and ballistic. During the ground-contact phase, we solve for the state and control trajectories by
\begin{eqnarray}
\max_{\vc{q}_{1:T}, \dot{\vc{q}}_{1:T},\boldsymbol{\tau}_{1:T}} & y_T + \frac{\dot{y}_T^2}{2g}, \\
\mathrm{subject\;to\;} & \ddot{\vc{q}}_t = \vc{S}(\boldsymbol{\tau}_t, \vc{q}_t, \dot{\vc{q}}_t),  \\
& -\boldsymbol{\epsilon} \preceq \vc{R}(\vc{q}_t, \dot{\vc{q}}_t, \boldsymbol{\tau}_t) \preceq \boldsymbol{\epsilon}, \label{eqn:torque-limits}\\
& \ddot{y}_t \geq -g, \mathrm{\;where\;} 1\leq t \leq T-1, \label{eqn:non-negative-force}\\
& \ddot{y}_T = -g,  \label{eqn:no-force} \\
& \vc{q}_1 = \bar{\vc{q}};\;\;\dot{\vc{q}}_1 = \bar{\dot{\vc{q}}},
\end{eqnarray}
where $y_T$ and $\dot{y}_T$ are the height and vertical velocity of the center-of-mass at the last frame of the ground-contact phase, and $g$ is the gravitational acceleration (9.8 m $\cdot$ s$^{-2}$). The contact is modelled by a revolute joint between the toe and the ground. Equation \ref{eqn:non-negative-force} enforces non-negative contact force in the vertical direction from frame $1$ to frame $T-1$. The contact force vanishes at the last frame of the ground-contact phase (Equation \ref{eqn:no-force}). The number of control points ($T$) used to represent the state and control trajectories is set to $200$ for the ground-contact phase, which is equivalent to the duration of one second. Since minimizing the metabolic energy is unimportant for the task of jumping as high as possible, we forgo the learned energy function in this example and only include the learned torque limits (Equation \ref{eqn:torque-limits}).

In the ballistic phase, the jumper is added two float-base DOFs and removed the revolute joint on the toe. The ballistic trajectory is solved using $200$ temporal nodes, which is equivalent to $0.5$ seconds. The initial state of the ballistic phase is defined by the final frame of the ground-contact phase. The final position is constrained to be fully upright and vertical. 
\begin{figure}
  \includegraphics[width=1.0\linewidth]{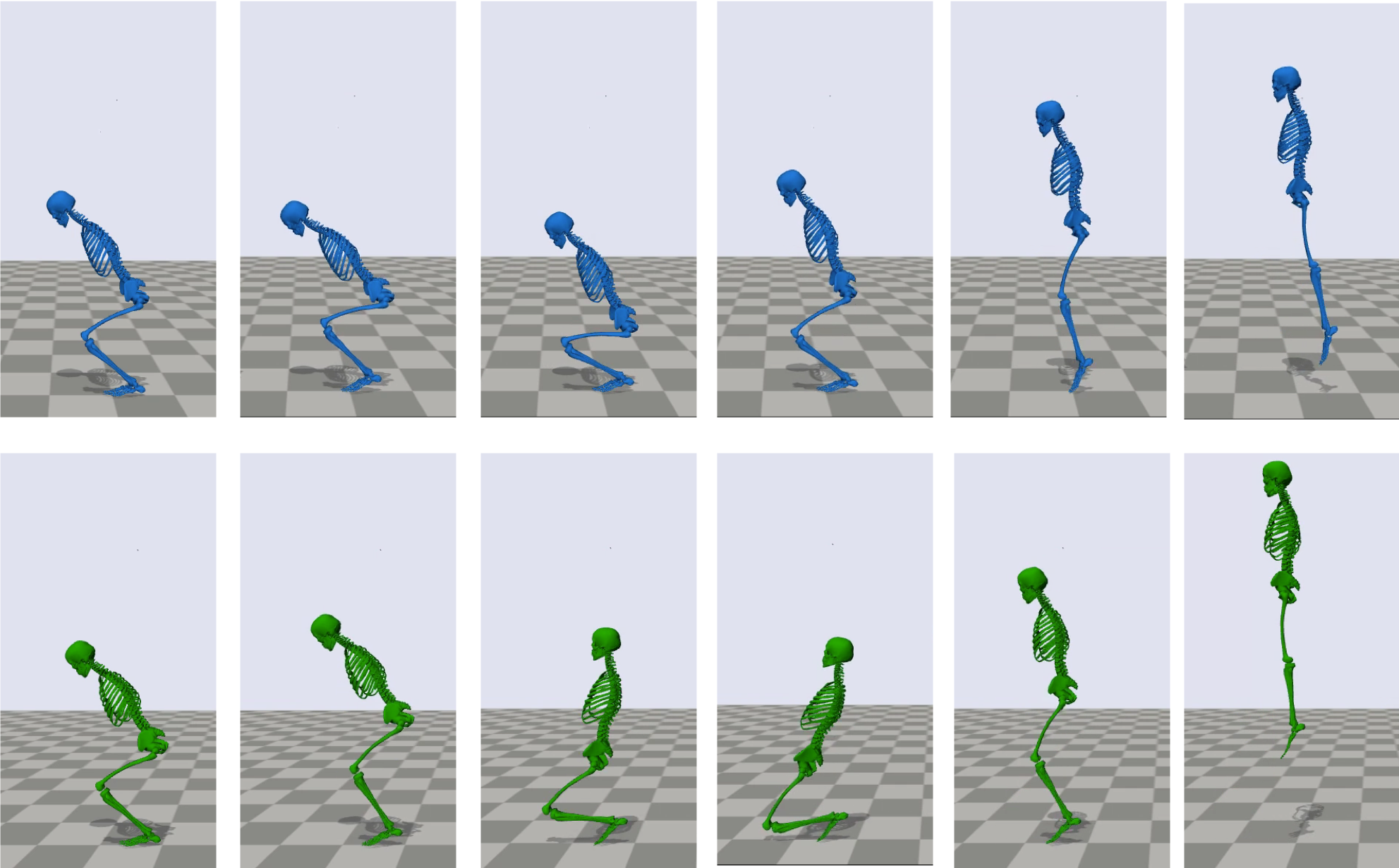}
  \caption{Top: Jumping motion using our learned torque limits. Bottom: Jumping motion using box torque limits.}
  \label{fig:jumping_OURS_BOX}
\end{figure}

Figure \ref{fig:jumping_OURS_BOX} compares our optimal motion trajectories against BOX. It is evident that the motion generated with conventional box torque limits (BOX) exhibits unnatural flexion. Our motion is visually similar to the motion generated by musculotendon model (MTU), as shown in the supplementary video. In terms of the maximal height, our method and MTU reach similar heights at $1.26$ m and $1.27$ m respectively, while BOX reaches $1.6$ m.

\begin{table}
\begin{center}
\captionsetup{justification=centering}
\caption{Performance of trajectory optimization. \\ Time solving for ballistic phase excluded in both experiments.}
\begin{tabular}{|c|c|c|c|c|c|c|}
\hline
\multirow{2}{*}{} & \multicolumn{2}{c|}{LR} & \multicolumn{2}{c|}{MTU} & \multicolumn{2}{c|}{BOX} \\ \cline{2-7} 
                  & time         & iter       & time        & iter       & time        & iter       \\ \hline
jump              & 733 s         & 159        & 1760 s       & 99         & 474 s        & 127        \\ \hline
max-swing          & 1115 s        & 280        & 7558 s       & 502        & 837 s        & 214       \\ \hline
\end{tabular}
\label{tab:performance}
\end{center}
\end{table}

Table \ref{tab:performance} compares the performance of LR, BOX, and MTU. MTU in this example takes fewer iterations but longer wall-clock time to solve the problem. This might be due to the fact that MTU formulates a much larger optimization problem in the muscle-activation space, but our inequality constraints on torque limits are more nonlinear than MTU's. As expected, BOX takes fewer iterations than LR due to simpler torque limit constraints, but we also note that each iteration of BOX and LR takes approximately the same wall-clock time, implying that the forward pass and back-propagation of trained $\vc{R}$ takes negligible time per iteration.

Though simple, this jumping example exposes the shortcomings of conventional box torque limits. Figure \ref{fig:jumping_OURS_BOX} Bottom shows that the character exploits hyper flexion of knee and ankle to create longer distance for the center-off-mass to accelerate upward. However, when the human knee and ankle are in such hyper-flexed position, they are unable to generate a large torque as computed by BOX. As an example, when the character is in the pose shown in the third image of the bottom of Figure \ref{fig:jumping_OURS_BOX}, the torque vector solved by BOX is $\vc{\tau} = [-50.9, 163.0, 45.8]$ (Nm), which does not violate the box torque limits. However, the closest valid torque vector according to our learned torque limits is $[-48.4. 102.2, 44.8]$, indicating that non-humanlike torques are used. We note that it is possible, through trial-and-error, to find more favorable box torque limits that result in better motions. We intentionally choose not to tweak the torque limits as they are usually task-dependent.


\newpage
\subsubsection{Swinging}
For the swinging task, we add the shoulder DOF and the arm segment to the 2D jumper. The implementation of shoulder is identical across our method and all baselines. That is, the shoulder uses joint-actuation with box torque limits $[0, 80]$ Nm. Starting from hanging vertically on a horizontal bar with zero velocity, we formulate two problems with different tasks. The first task is to generate momentum by swinging back and forth, such that the flight distance is maximized after the character releases the bar. The second task is to reach a fixed distance after the character releases the bar while minimizing the effort. The motion is solved in two phases: swing and ballistic. We describe the formulation of each task below.

\textbf{Maximum flight.} During the swing phase, we solve the state and control trajectories by: 
\begin{eqnarray*}
\max_{\vc{q}_{1:T}, \dot{\vc{q}}_{1:T},\boldsymbol{\tau}_{1:T}} & \dot{x}_T \cdot \dot{y}_T,  \\
\mathrm{subject\;to\;} & \ddot{\vc{q}}_t = \vc{S}(\boldsymbol{\tau}_t, \vc{q}_t, \dot{\vc{q}}_t),  \\
& -\boldsymbol{\epsilon} \preceq \vc{R}(\vc{q}_t, \dot{\vc{q}}_t, \boldsymbol{\tau}_t) \preceq \boldsymbol{\epsilon} , \\
& \vc{q}_1 = \bar{\vc{q}};\;\;\dot{\vc{q}}_1 = \bar{\dot{\vc{q}}},
\end{eqnarray*}
where $\dot{x}_T$ and $\dot{y}_T$ are the horizontal and vertical velocities of the center-of-mass at the last frame of the swing phase. We use $200$ control points to represent the trajectory of $3$ seconds for this example.

The ballistic phase is similar to the jumping example. A 2D float-base is added to the model and the revolute joint between hand and the bar is removed. The ballistic trajectory is solved for 60 control points with a fixed time horizon. The initial state is defined by the final frame of the swing phase. For the final frame, we calculate the center-of-mass location at the end of ballistic phase and set a nearby location to be the target of the hand. The motion is as if the model would jump and grasp a next bar. The target for BOX is $2$ m horizontally from the first bar and $1.5$ m for MTU and our method.

Similar to jumping, we observe that our method generates similar motion to MTU (Figure \ref{fig:swinging_max}). MTU reaches a final $\dot{x}_T $ of 2.87 m/s, $\dot{y}_T$ of 2.39 m/s. Our method reaches a final $\dot{x}_T$ of 2.86 m/s, $\dot{y}_T$ of 2.3 m/s. On the other hand, the motion produced by BOX overly flexes and extends the knee (see supplementary video), which results in an unrealistically large take-off velocity ($\dot{x}_T$ = $3.74$ m/s, $\dot{y}_T$ = $3.56$ m/s).

Comparing to MTU, the performance gain of our method is more evident in this example with LR taking fewer iterations than MTU, possibly due to solving the problem in a lower-dimensional space (Table \ref{tab:performance}). We will see even more significant performance gain when applying our method to examples in deep reinforcement learning.

\textbf{Fixed distance with minimum energy.} This example evaluates our learned energy function by comparing LR+LE with the standard effort function that penalizes the sum of absolute value of torques (LR). The state and control trajectories in the swing phase are generated by:
\begin{eqnarray*}
\min_{\vc{q}_{1:T}, \dot{\vc{q}}_{1:T},\boldsymbol{\tau}_{1:T}} & \sum_{t=1}^T E(\vc{q}_t, \dot{\vc{q}}_t, \boldsymbol{\tau}_t),  \\
\mathrm{subject\;to\;} & \dot{x}_T \cdot \dot{y}_T = 4.0, \\
& \ddot{\vc{q}}_t = \vc{S}(\boldsymbol{\tau}_t, \vc{q}_t, \dot{\vc{q}}_t),  \\
& -\boldsymbol{\epsilon} \preceq \vc{R}(\vc{q}_t, \dot{\vc{q}}_t, \boldsymbol{\tau}_t) \preceq \boldsymbol{\epsilon}, \\
& \vc{q}_1 = \bar{\vc{q}};\;\;\dot{\vc{q}}_1 = \bar{\dot{\vc{q}}},
\end{eqnarray*}
where the constraint $\dot{x}_T \cdot \dot{y}_T = 4$ on the final state enforces a fixed flight distance after the character releases the bar. We use $350$ control points to represent the trajectory of $5$ seconds for this example.

During the ballistic phase, we formulate an optimization similar to the case of maximum flight, except that the final position of the hand in the air is constrained to a relative horizontal distance of $0.75$ m to the first bar.

Although both LR+LE and LR reach the same distance, we observe differences in the state and torque trajectories due to the different energy functions (see supplementary video). One noticeable difference is that LR tends to flex the knee while LR+LE lets the shin segment swing more passively. 

\subsection{Policy learning}
\label{sec:policy-learning}
We demonstrate that our method can be used in conjunction with policy learning algorithms to learn locomotion policies without the use of motion data. Due to the stochastic nature of policy learning, there is no theoretical equivalence between the learning problems formulated in the joint-actuation space and muscle-actuation space, as is the case with trajectory optimization. However, the evaluation in this section still demonstrates the benefits provided by our method to the state-of-the-art policy learning.

We build our experiments upon previous work \cite{Yu:2018} and its open-source implementation. By providing scheduled, decremented assistive forces and penalizing asymmetric actions, Yu \etal showed that low-energy, symmetric locomotion policies can be learned without motion data or morphology-specific knowledge. However, the authors noted that careful tuning of the character model is needed for generating natural motion. In particular, the parameters for the range of action (\ie torque limits) and the joint stiffness and damping on each joint play an important role on the quality of resultant motion. In contrast, our method replaces manually tuned joint spring-dampers and relies on the learned energy function to account for the effect of passive elements in the human musculoskeletal system. In addition, the learned, state-dependent torque limits further eliminate the need to tune the range of action space for every task.


We use an under-actuated 3D human model which consists of $13$ segments and $29$ DOFs. The upper-body DOFs are all actuated by joint actuators with box torque limits, while the lower-body DOFs are implemented differently in different methods---MTU actuates the lower-body DOFs using muscles, LR+LE and LR use joint actuators with learned torque limits, and BOX uses joint actuators but with box torque limits (200 Nm for the flexion DOFs of hip, knee and ankle, 60 Nm for the other two DOFs of hip). No joint springs are used and the joint damping is set to a small, uniform value across all lower-limb DOFs for numerical stability. 

Directly learning a DRL policy on a complex muscle-based model (\ie MTU) is computationally too costly---each learning iteration of MTU takes about $7.5$ minutes comparing to $30$ seconds for LR+LE, LR, or BOX. As such, we use AMTU as an improved MTU baseline. AMTU trains a regressor to approximate the forward muscle dynamics function by mapping a muscle activation vector in a given state to the corresponding torque vector. As an indicator that AMTU is indeed a reasonable approximator of MTU, we trained a walking policy using the regressor of AMTU and tested the policy using the exact muscle dynamics in MTU. The results show that the character using that policy can successfully walk with nearly identical motion to the one tested with AMTU. As another validation, when training the walking policy, the first $300$ iterations of learning curves were nearly identical between MTU and AMTU. It took $36$ hours for MTU to run $300$ iterations, while it only took $2.5$ hours for AMTU.

Our formulation has the same state space and action space as those described in Yu \etal \shortcite{Yu:2018}. A state $\hat{\vc{s}}$ includes $\vc{q}$, $\dot{\vc{q}}$, two binary bits to indicate the contact state of the feet, and the target velocity of the gait. An action $\hat{\vc{a}}$ is simply the joint torque generated by the actuators of the character. Our reward function follows a similar definition as in Yu \etal \shortcite{Yu:2018}, with the modifications introduced in \ref{sec:RL-implement} if using LR+LE:
\begin{equation}
r(\hat{\vc{s}},\hat{\vc{a}}) =  4.5 E_{v}(\hat{\vc{s}}) + E_u(\hat{\vc{s}}) + 3 E_l(\hat{\vc{s}}) + 9 + w_e E_{e}(\hat{\vc{s}}, \hat{\vc{a}}).
\end{equation}
Here $E_e$ is the effort term, and other terms $E_v, E_u, E_l$ are the same as in Yu \etal \shortcite{Yu:2018} to maintain balance and move forward without specifying the kinematics or style of the gait.


Since LR+LE uses the learned $E$ as the effort term while LR, BOX and AMTU use the sum of normalized torques, the choice of the effort weight, $w_e$, can influence the results differently for different methods. We address this potential bias by training and comparing multiple policies across a range of $w_e$ for each method, instead of arbitrarily determining a single value for $w_e$. 



\subsubsection{Walking}
\begin{figure}
  \includegraphics[width=1.0\linewidth]{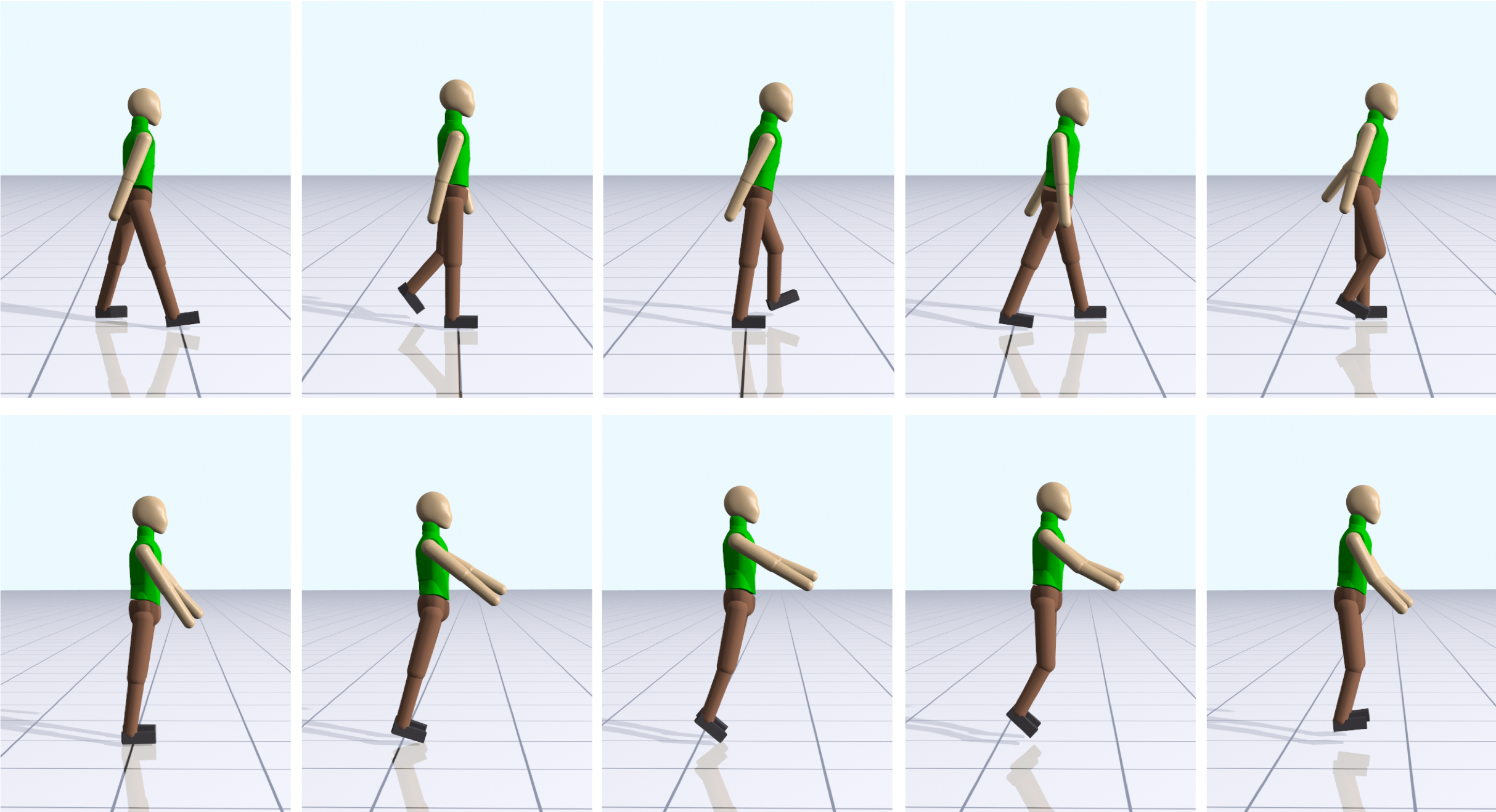}
  \caption{Top: Walking motion from a policy trained with a larger effort penalty weight. Bottom: Hopping motion from a policy trained with a smaller effort penalty weight. The two training regimes are identical otherwise (LR+LE).}
  \label{fig:walking_hopping_OURS}
\end{figure}

The following results are best evaluated by viewing the supplementary video. Our results show that without tuning of joint spring coefficients or joint torque limits, the policy trained by BOX produces gait that generates large angular movements about the yaw-axis and high-frequency ankle motion that requires unnaturally large joint torques. When enforced with our learned torque limits (LR), the agent learns to only use feasible torques to walk. If we further replace the effort term with our learned metabolic energy function (LR+LE), the agent learns to lower the gait frequency and take larger stride, as well as reducing unnecessary angular movements about the yaw axis (Figure \ref{fig:walking_hopping_OURS} Top).

One thing worth noting is that the minimalist reward function proposed by Yu \etal can also lead to hopping, depending on the energy penalty weights, $w_e$. With a relatively higher $w_e$, BOX learns a walking gait with unnaturally fast ankle movement, while learning a hopping gait with similarly fast ankle movement when $w_e$ is lower. AMTU consistently learns a walking gait in the range of $w_e$. In terms of our method, when $w_e$ is high, our walking gait (LR) is similar to AMTU and more natural looking than BOX. When $w_e$ is lower, our method produces a natural hopping gait (Figure \ref{fig:walking_hopping_OURS} Bottom). All methods fail to learn a successful walking policy when $w_e$ is too high. Note that we are not able to train AMTU with the sum-of-activation energy term; in our experiments AMTU only works with sum-of-torque energy formulation.


Since the computation time for each iteration is similar among BOX, LR+LE, LR, AMTU, we can directly compare the number of iterations to evaluate the performance of each method. In our experiments, BOX takes slightly fewer iterations than LR and LR+LE. For AMTU, the number of iterations varies with different $w_e$ and different random seeds, ranging between 20 to 100\% more than our method. Directly using MTU takes $15$ times longer to compute each iteration than our method.

\begin{figure}
  \includegraphics[width=0.9\linewidth]{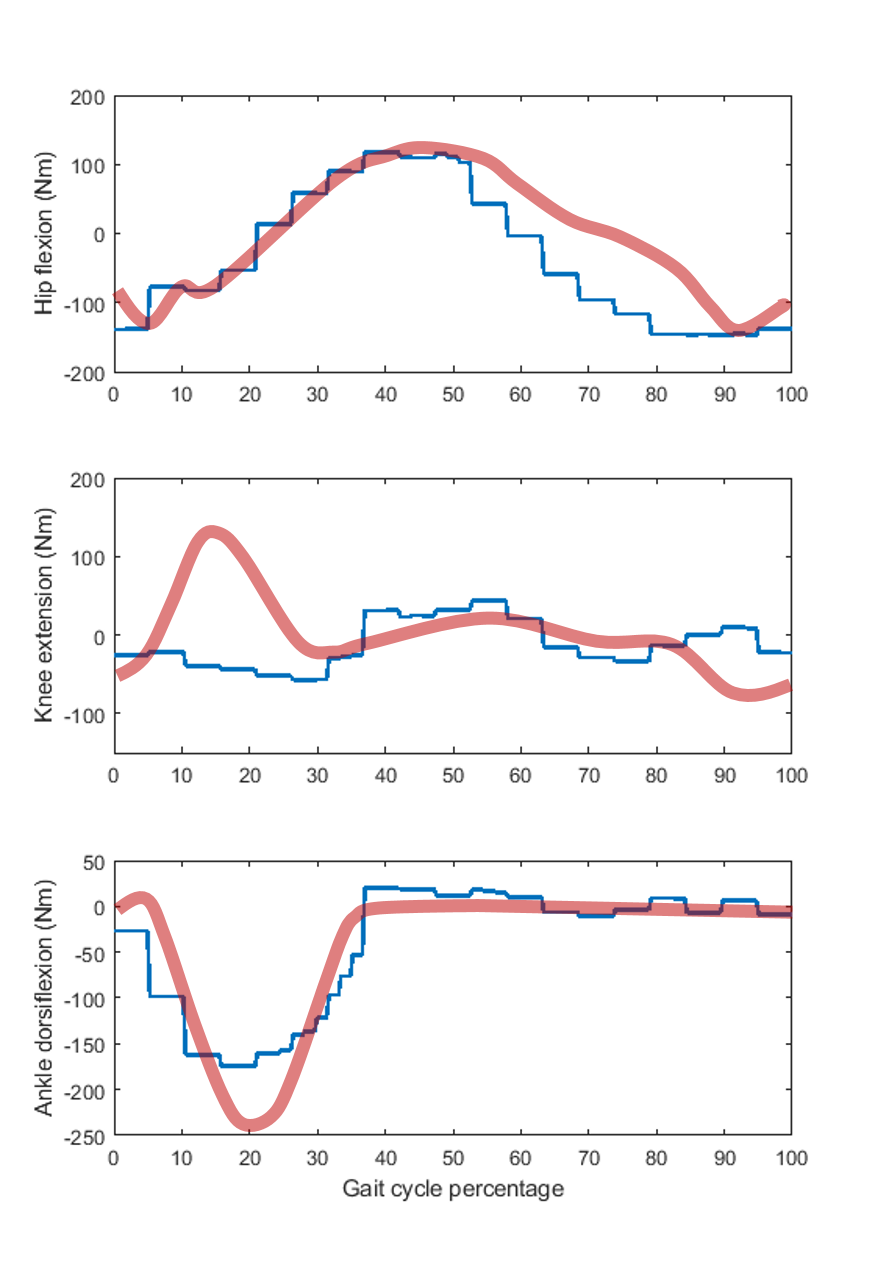}
  \caption{Torque patterns generated by our running policy (blue) comparing with the human data (red).}
  \label{fig:torque_traj_run}
\end{figure}

Similar to \cite{Yu:2018}, we are interested in learning with the minimalist approach and intentionally restrain from fine-tuning the reward function for improving the style of the motion. For example, the arm motion can be largely improved if we include additional target poses in the reward function. Without any specification of a desired gait in the reward function, the arm has relatively low mass and thus its movement has little impact on the overall dynamics. We notice that the arm motion generally varies by random seeds (see supplementary video).


\subsubsection{Running}
To train a running policy, we simply increase the target speed in the reward function and keep everything else the same as for training the walking policy. We again compare our method against AMTU and BOX across a range of $w_e$. Unlike learning a walking policy, AMTU is not able to learn any successful policy for the entire range of $w_e$. BOX consistently learns a high-frequency hopping policy in the range of $w_e$. For our method, both LR+LE and LR successfully learn a running gait. The difference in the resulting motion is negligible between the two, as minimal use of energy plays a less important role for highly dynamic motion, such as running. 

We compare the torque patterns generated by our learned policy with those reported in the biomechanics literature \cite{Wang:2012} (Figure \ref{fig:torque_traj_run}). Note that our torque patterns are piecewise-constant due to the low control frequency of the policy (33 Hz). Figure \ref{fig:torque_traj_run} shows that our policy produces hip and ankle torque patterns similar to the human data. However, our policy does not learn to exert large torque to extend the knee at the beginning of the gait cycle. In accordance with Ackermann \etal \shortcite{ackermann2010optimality}, simulated gait based on energy minimization exhibits less knee flexion during stance as observed in human movement. This gait pattern requires smaller knee extension torques. Biomechanics researchers have suggested that gait with near-zero joint torques is unstable and that taking uncertainty into account would lead to more optimal movements with larger knee flexion during stance and larger knee extension torques as consequence \cite{koelewijn2018metabolic}.

\begin{figure}[t!]
    \centering
    \begin{subfigure}[t]{0.25\textwidth}
        \includegraphics[height=1.3in]{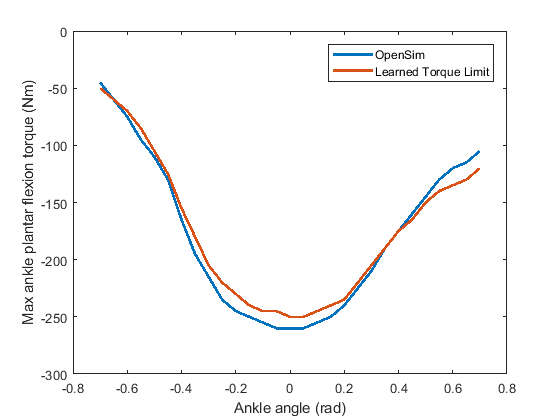}
    \end{subfigure}%
    \begin{subfigure}[t]{0.25\textwidth}
        \includegraphics[height=1.3in]{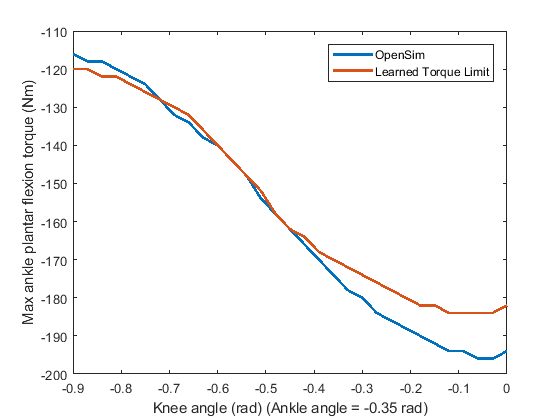}

    \end{subfigure}
    \caption{Ankle torque limit at different poses. Left: The ankle torque limit is lower when the ankle is in a flexed or extended position. Right: The ankle torque limit is lower when the knee is in a flexed position.}
    \label{fig:R-visualization}
\end{figure}

\begin{figure}[t!]
    \centering
    \begin{subfigure}[t]{0.25\textwidth}
        \includegraphics[height=1.3in]{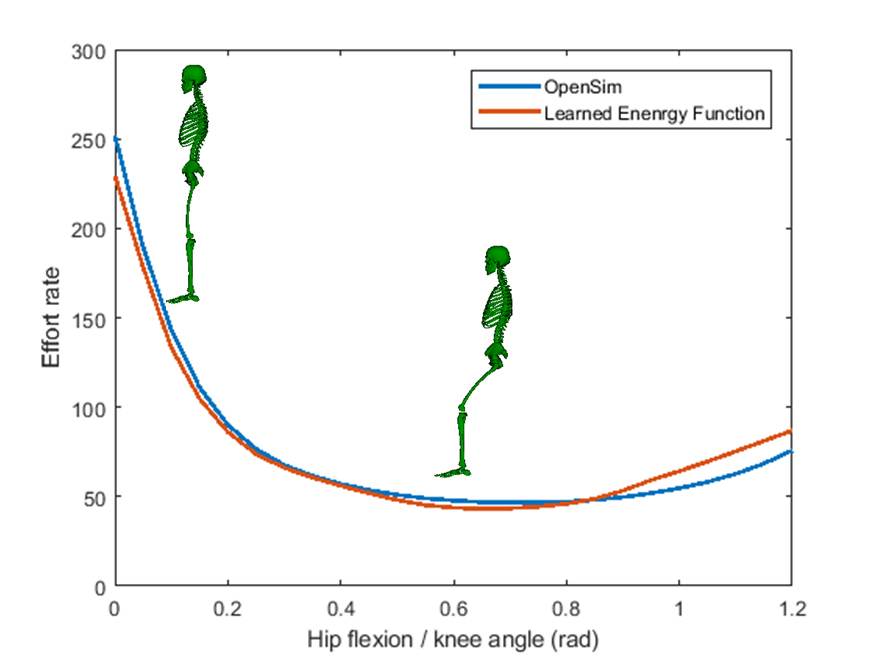}
    \end{subfigure}%
    \begin{subfigure}[t]{0.25\textwidth}
        \includegraphics[height=1.3in]{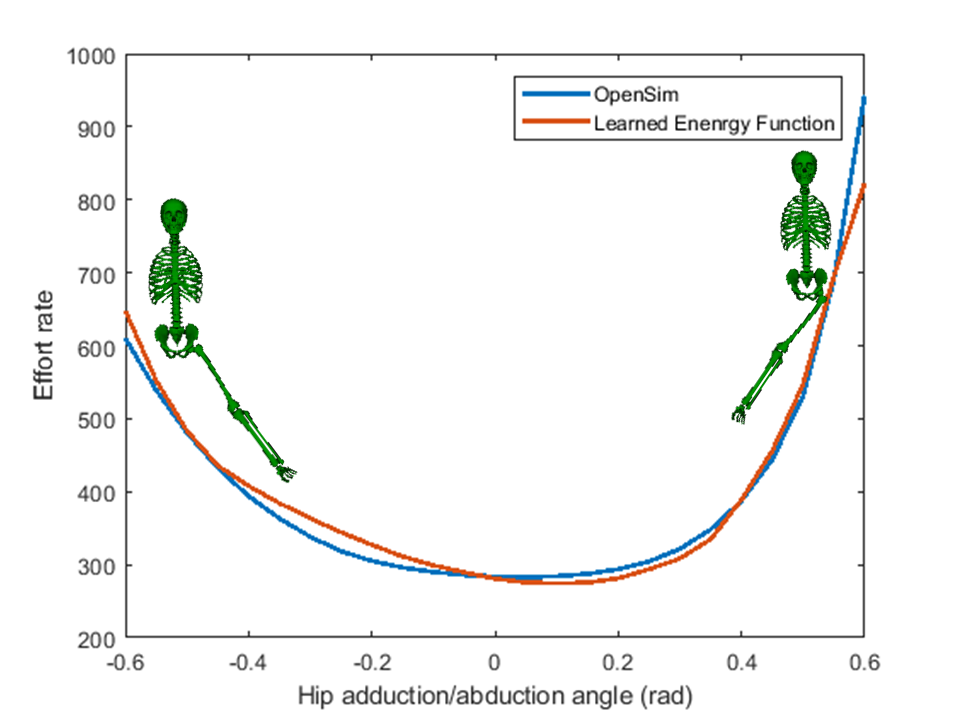}

    \end{subfigure}
    \caption{Effort rate required to generate 80 Nm of torque at hip when the character is in different poses. Left: Generating torque to extend the hip is easier when the character's hip and knee flex. Right: Generating torque to flex the hip is harder when the character's hip is adducted or abducted. }
    \label{fig:E-visualization}
\end{figure}

\subsection{Visualizing learned torque limits and energy function}
Figure \ref{fig:R-visualization} visualizes the learned torque limits for different states. We plot the upper bound of ankle torque for a range of ankle angles and fix other variables of the state at zero value (Figure \ref{fig:R-visualization} Left). The maximal torque is lower when the ankle is more flexed or extended, consistent with the findings in biomechanics literature. The torque limits also depend on the state of other DOFs. Figure \ref{fig:R-visualization} Right shows that the more knee flexes the less torque ankle is allowed to generate. 

We also visualize the learned energy function over different states. Figure \ref{fig:E-visualization} shows the amount of effort required to generate $80$ Nm of torque at the hip when the character is in various poses. When both hip and knee flex as opposed to be at the zero position, it takes less effort for the hip to generate torque (Figure \ref{fig:E-visualization} Left). On the other hand, it the hip adducts or abducts, it is harder for the hip to generate torque (Figure \ref{fig:E-visualization} Right).
\section{Discussion}
\label{sec:discussion}
While our work enables some muscle-like behaviors without explicitly modeling muscles, we have made a few important assumptions. First, our model does not include tendon compliance thus ignoring contraction dynamics in modeling MTUs. Tendon compliance does play an important role when computing metabolic energy consumption \cite{uchida2016stretching}. However, it is inconclusive as to what extent the tendon compliance affects patterns of submaximal movements \cite{anderson2001static, lin2012comparison}. For few MTUs with long tendons, such as the Soleus, tendon compliance is generally believed more important, especially in tasks like sprinting. Incorporating tendon compliance can be an important research direction in the future.

Another simplification in our model is the exclusion of activation dynamics, which limits the rate of muscle activation. While the muscle response time is generally short---$10$ ms for activation and $40$ ms for deactivation \cite{millard2013flexing}---it is possible that controlling neural excitation rather than muscle activation plays a crucial role for certain tasks. One possible future direction is to build a torque-rate-limit function to account for activation dynamics.

As mentioned in Section \ref{sec:related-work}, many other energy expenditure formulae exist. It is possible that the energy function used in this work does not reflect the behaviors of human musculoskeletal system for certain tasks. Since our proposed framework is sufficiently general, one can experiment with other energy formulation such as fatigue or robustness. 

Our current method is not able to model muscle co-contraction---simultaneous contraction of the agonist and the antagonist muscles around a joint to hold a static position. This is because muscle co-contraction directly violates our assumption that human generates minimal-effort activation pattern for a given torque. As a future direction, one can consider augmenting additional time-varying internal muscle states to model muscle co-contraction on torque actuation level.


Both LR and LR+LE can learn walking and running policies consistently, but AMTU fails to learn a running policy. For walking, AMTU fails on learning a policy with the energy function that minimizes muscle activation, but is able to learn successful policies when a simpler sum-of-torque energy function is used. One conjecture is that by optimizing policy in the joint-actuation space, we are effectively searching in the space of minimal-effort muscle activations, a subspace of the high-dimensional activation space, making the policy optimization more tractable.

Our intention to implement AMTU was to provide a more practical baseline for the muscle-based model when solving a policy learning problem. Directly applying MTU to our learning problems would take days of computation time and the learning outcome might still be inconclusive. Although AMTU does not reduce the difficulty of learning in the high-dimensional action space, it at least makes the training process more tractable by speeding up the computation by $15$ times. If a policy learning must be solved in the muscle-actuation space, one can consider training an initial policy using AMTU and refine it with MTU if necessary. One can as well consider applying AMTU to trajectory optimization problems to speed up MTU in each iteration, though a save in the number of total iterations would be unlikely.

\section{Conclusions}
We present a new technique to transform an optimal control problem formulated in the muscle-actuation space to an equivalent problem in the joint-actuation space. By solving the equivalent problem in the joint-actuation space, we can generate human-like motion comparable to those generated by musculotendon models, while retaining the benefit of simple modeling and fast computation offered by joint-actuation models. Comparing to the commonly used box torque limits, our method produces more human-like movements and torque patterns, as well as eliminates the need to tune the torque limits for each specific task. Our method lends itself well to both trajectory optimization and policy learning using deep reinforcement learning approach, making the control problems more tractable by optimizing in a lower-dimensional space. 

\begin{acks}
The authors would like to thank Maarten Afschrift, Visak CV Kumar, Xia Wu, Yujia Xie, Wenhao Yu and Yunbo Zhang for insightful discussions and technical assistance. The authors would also like to thank the anonymous reviewers for helpful suggestions. The work is supported by NSF Award No. 1514258 and the 2017 Samsung GRO Program Award.
\end{acks}

\bibliographystyle{ACM-Reference-Format}
\bibliography{sample-bibliography}

\end{document}